**Giant quadratic magneto-optical response of thin YIG films for sensitive magnetometric experiments**

E. Schmoranzerová[1]*, T. Ostatnický[1]*, J. Kimák[1], D. Kriegner[2,3], H. Reichlová[2,3], R. Schlitz[3], A. Baďura[1], Z. Šobáň[2], M. Münzenberg[4], G. Jakob[5], E.-J. Guo[5], M. Kläui[5], P. Němec[1]

[1]Faculty of Mathematics and Physics, Charles University, Prague, 12116, Czech Republic

[2]Institute of Physics ASCR v.v.i , Prague, 162 53, Czech Republic

[3]Technical University Dresden, 01062 Dresden, Germany

[4]Institute of Physics, Ernst-Moritz-Arndt University, 17489, Greifswald, Germany

[5]Institute of Physics, Johannes Gutenberg University Mainz, 55099 Mainz, Germany

**Abstract**

We report on observation of a magneto-optical effect quadratic in magnetization (Cotton-Mouton effect) in 50 nm thick layer of Yttrium-Iron Garnet (YIG). By a combined theoretical and experimental approach, we managed to quantify both linear and quadratic magneto-optical effects. We show that the quadratic magneto-optical signal in the thin YIG film can exceed the linear magneto-optical response, reaching values of 450 rad that are comparable with Heusler alloys or ferromagnetic semiconductors. Furthermore, we demonstrate that a proper choice of experimental conditions, particularly with respect to the wavelength, is crucial for optimization of the quadratic magneto-optical effect for magnetometry measurement.

I. **Introduction**

Yttrium Iron Garnet ($Y_3Fe_5O_{12}$, YIG) is a prototype ferrimagnetic insulator which represents one of the key systems for modern spintronic applications [1]. It has been thoroughly studied in the last decades owing to its special properties, such as low Gilbert damping [2-4] and high spin pumping efficiency [5-7]. YIG has played a crucial role in fundamental spintronics experiments, revealing spin Hall magneto-resistance [9-10] or spin-Seebeck effect [11-13].

Many of the above-mentioned spintronic phenomena rely on high-quality ultra-thin YIG films and on detection of small changes in magnetization therein. However, YIG is a complex magnetic system with 200 $\mu_B$ magnetic moments per unit cell. Magnetic properties of the few monolayer systems used in spintronics are then vulnerable to small structural changes and relatively difficult to characterize and control [14-17]. Moreover, reliability of conventional magnetometry tools such as Superconducting Quantum Interference Device (SQUID) or Vibrating Sample Magnetometry (VSM) is limited by the large paramagnetic background and unavoidable impurity content of the gadolinium-gallium- garnet that is commonly used as a substrate for the thin YIG layers. Direct use of magneto-transport methods for magnetic characterization is naturally prevented by the small electric conductivity of the insulating YIG. They can be utilized only indirectly in multilayers of YIG/heavy metal, via Spin Hall Magnetoresistance in the metallic layer [17].

In contrast, optical interactions are not governed by DC conductivity of the material. Magneto-optics therefore provides a natural tool for detection of magnetic state of ferrimagnetic insulators, and YIG in particular has an extremely strong magneto-optical response that can be easily modified by doping [18]. Magneto-optical (MO) response of a material manifests generally as a change of polarization state of a transmitted or reflected light [18] usually detected in a form of a rotation of polarization plane of a linearly polarized light. Similar to the magneto-transport effects, MO effects with different symmetries with respect to magnetization (**M**) can occur. With certain limitations [19], an optical analogy to the anomalous Hall effect (AHE) linear in **M** represents the Faraday effect in transmission geometry or the Kerr effect in reflection. For the anisotropic magnetoresistance (AMR) quadratic in **M**, the corresponding MO effect is magnetic linear dichroism (MLD) [19]. As the terminology is ambiguous in magneto-optics, MLD, Q-MOKE, Voigt or Cotton-Moutton effect, which are sometimes used, refer all to the same phenomenon in different experimental geometries. In this paper, we use the name Cotton-Moutton effect (CME) for the rotation of polarization plane in transmission geometry, consistently with previous works on YIG [20-21].

The quadratic MO effects scale with the square of the magnetization magnitude and their symmetry is given by an axis parallel to the magnetization vector. As such, they are generally weaker than the linear MO response [18]. However, there are significant advantages over the linear magneto-optics that make them favourable in MO magnetometry. The even symmetry with respect to the local magnetization enables to observe these effects in systems with no net magnetic moment, such as collinear antiferromagnets, as the contributions from different sublattices do not cancel out [22]. Quadratic MO effects are sensitive to the angle between magnetization and the polarization plane [23], similarly to the way the AMR is sensitive to the angle between the electric current and the magnetization [19], which enables to trace all the in-plane components of magnetization vector simultaneously in one experiment [23-26] There is, however, a key advantage of the MO approach. The optical polarization can be set easily without fabrication of additional structures, unlike the current direction in the case of AMR, which is given by the electrical contact geometry. Variation of the probe polarization can then provide information about magneto-crystalline anisotropies [25] without modification of sample properties by litography-induced changes that are inherent to the methods based on electron transport.

In certain classes of materials for compunds with significant spin-orbit coupling, such as Heusler alloys [27], ferromagnetic semiconductors [23,28] or some collinear antiferromagnets [22], the quadratic MO response can be strongly enhanced. It has found important applications in static and dynamic MO magnetometry [28-29], helping to reveal novel physical phenomena such as optical spin transfer [30] and spin orbit torques [31]. In contrast, in ferrimagnetic insulators the quadratic MO effects have been vastly neglected so far. The first pioneering experiments have revealed the potential of the quadratic magneto-optics in YIG to visualize stress waves [32] or current-induced spin-orbit torque [21], and the inverse quadratic Kerr effect has been even identified as a trigger mechanism for ultrafast magnetization dynamics in thin YIG films [21]. However, no optimization with respect to the size of the amplitude of MO effects was performed in these works in terms of its amplitude, spectrum, dependence on the angle of incidence or initial polarization. In 1970s and 80s, limited studies were performed in the field of magneto-optical spectroscopy on bulk YIG crystals, demonstrating magnetic linear birefringence [33] or dichroism [34-35] of YIG crystals doped by rare-earths and metals, and on terbium-gallium-garnet at cryogenic temperature [36]. However, to our knowledge, no experiments aiming at understanding the details of the quadratic MO response in thin films of pure, undoped YIG have been performed so far.

In this paper, we report on the observation of a giant CME of 50 nm thin epitaxial film of pure YIG, which can even exceed the amplitude of the linear Faraday effect. Using a combined experimental and theoretical approach, we quantify the size of CME with respect to various external parameters, such as wavelength, temperature or angle of incidence. This is a key prerequisite for the quadratic magneto-optics optimization for magnetometric applications. The potential of CME for magnetometry is demonstrated on identification of magnetic anisotropy of the thin YIG film directly from the detected MO signals.

## II. Experimental details and sample characterization

We used monocrystalline 50 nm thick film of yttrium iron garnet prepared by pulsed laser deposition (PLD) on (111)-oriented gallium gadolinium garnet (GGG). Details of the growth procedure can be found in Ref. 37. Since the thin YIG layers are prone to growth defects and strain inhomogeneities [14],[16], the samples were carefully characterized by X-ray diffraction. From the reciprocal space maps (RSM) around the YIG and GGG 444 and 642 Bragg peak we find that any diffraction signal of the film is aligned with the one of the substrate along the in-plane momentum transfer [see Fig. 1(a) and (b)]. The RSMs therefore show the pseudomorphic growth of the YIG film, with its in-plane lattice parameter equal to the substrate lattice parameter measured to be 12.385 Å, close to previously published values for GGG substrates [38]. Along the [111] direction we find Laue thickness oscillations indicating the high crystalline quality of the YIG film. In order to analyze the out of plane lattice parameter, a cross-section along the [111] direction was extracted from the 444 and 642 Bragg peaks and modelled using a dynamical diffraction model [see Fig. 1(c)]. We used the rhombohedrally distorted structure of YIG with a= 12.379 Å and rhombohedral angle α = (90.05  0.02) deg as an input for the model, parameters that are similar to Ref. [14]. A weak in-plane tensile strain occurs due to the lattice mismatch, but the resulting distortion of only 0.05 deg is unlikely to affect the magnetic properties of the layer in a significant way.

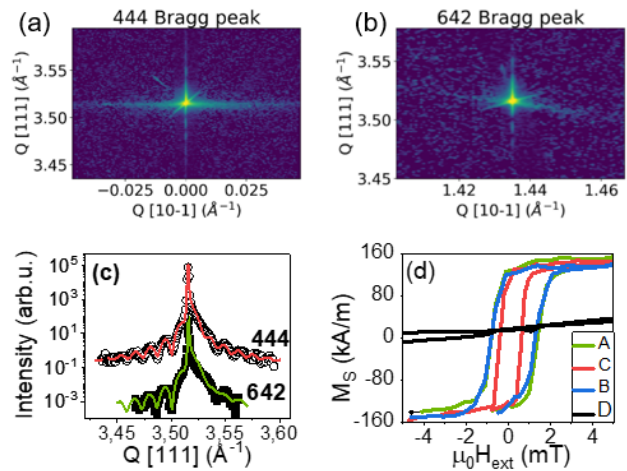

Fig. 1: Structural and magnetic characterization of thin YIG film. (a), (b) Reciprocal space maps (RSM) taken on 444 and 642 Brag peaks of YIG at room temperature. (c) Cross sections of RSM data along [111] crystallographic directions (points) modelled by dynamical diffraction model (line) with lattice parameter a = 12,379 Å and distortion angle α= (90,05  0.02) deg. (d) Magnetic hysteresis loops measured by SQUID magnetometry in three in-plane crystallographic directions that are denoted as: A [2-1-1], B[01-1] and C (diagonal), and in the out-of-plane direction D [111] at temperature of 50 K. T

The magnetic properties of the YIG film are established using SQUID magnetometry. An example of hysteresis loops recorded at 50 K with external magnetic field applied along different crystallographic directions is shown on Fig. 1(d). Clearly, the sample is in-plane magnetized, with a coercive field of $H_c$ = 18 Oe. Note that there is a small difference in $H_c$ between the two crystallographic directions denoted as *A* and *C*, indicating the presence of an in-plane magnetic anisotropy but its quantification based on our SQUID measurement is prevented by a large error caused by the paramagnetic background of the GGG substrate. For further details on SQUID measurements see Supplementary material, Fig. (S1). From these measurements, the room-temperature value $M_s$ = 96 kA/m was extracted. The saturation magnetization is lower than that of a bulk crystal $M_{s,bulk}$ = 143 kA/m [39] but in very good agreement with previously reported values for PLD-grown ultra-thin YIG layers [17], confirming the good quality of our YIG film.

Magneto-optical measurements on the YIG sample were performed by a home-made vectoral magneto-optical magnetometer, schematically shown in Fig. 2 (a). For the majority of our experiments we used a CW solid-state laser (Match Box series, Integrated Optics ltd.) with a fixed wavelength of 403 nm as a light source. The CW laser was replaced by the second-harmonics output of a tunable titan-sapphire pulse laser (model Mai Tai, Spectra Physics,) to gain wider spectral range of λ = 390-440 nm for the wavelength-dependence measurement. The light was incident on the sample either under an angle $\theta_i$ = 3° (near normal incidence), or $\theta_i$ = 45°, as indicated in Fig. 2(c). The linear polarization of the incident light in both cases was set by a polarizer and a half-waveplate, and the polarization state of the light transmitted through the sample was analyzed by a differential detection scheme (optical bridge) in combination with a phase-sensitive (lock-in) detection [40].

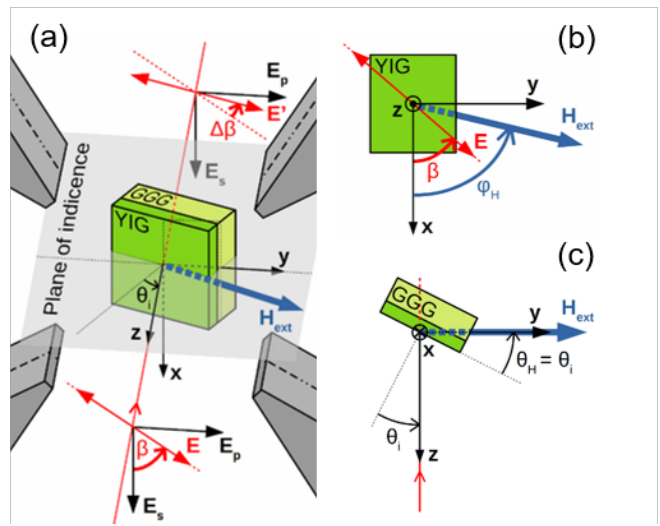

Fig. 2: (a) Schematics of the experimental setup for magneto-optical magnetometry. Linearly polarized light with a polarization **E** oriented at an angle $\beta$ with respect to the TM polarization mode ($E_s$) is incident on the sample, which is oriented under an angle $\theta_i$ with respect to the plane in which the magnetic field was applied. After being transmitted through the sample, the light polarization plane is rotated by an angle $\Delta\beta$. The sample is subject to an external magnetic field **H**$_{ext}$, applied in an arbitrary direction, with the corresponding spherical angles of the **H**$_{ext}$ vector shown in (b) plane view projection (azimuthal angle $\phi_H$) and (c) side view projection (polar angle $\theta_H$) of the experiment geometry.

The sample itself was mounted in a closed-cycle cryostat (ARS systems) to enable the temperature variation in a range of T = 20 K – 300 K. The cryostat was placed between pole stages of a custom-made two-dimensional (2D) electromagnet where the external magnetic field of up to $\mu_0 H_{ext}$ = 205 mT could be applied in an arbitrary direction in the plane perpendicular to the optical beam axis. The (spherical) coordinate system for $H_{ext}$ is given in Fig. 2(b), (c). Note that the polar angle $\theta_H$ is defined from the sample normal, and is equivalent to the angle of incidence of the incoming light $\theta_i$. Utilization of the 2D-electromagnet allowed for two different approaches in our experiments. Firstly, a standard magneto-optical magnetometry was used, where the magnitude of |**H**$_{ext}$| in a fixed direction [01-1] is varied, and the resulting hysteresis loops are recorded. Comparing the measured hysteresis loops for different orientations of the light polarization with our analytical model allowed for determination of the motion of magnetization during the magnetic field sweeps, as further discussed in the "Theory" section. Analysis of the full polarization dependence of the hysteresis loops also enabled us to separate the contributions of the linear Faraday effect (LFE) and the quadratic CME to the overall MO signals, and to extract the corresponding amplitudes (coefficients) of the CME and LFE effects.

However, this method of determining the MO coefficients was inefficient and burdened by a relatively large error resulting from the complicated way of extracting the MO effects that required full light polarization dependence of the hysteresis loops. For further systematic study of the CME effect we, therefore, implemented ROT-MOKE experiment, where the external magnetic field (**H**$_{ext}$) of a fixed magnitude of 205 mT was rotated in in the plane from $\phi_H$ = 0° to 360° (see Fig. 2), and the resulting MO signal was recorded as a function of $\phi_H$ [24-25], with the polarization of the light kept fixed to the fundamental TE (s-) mode. Here, **H**$_{ext}$ was large enough to saturate magnetization of the YIG film, which then exactly follows the field direction during its rotation. We can therefore neglect the effect of magnetic anisotropy and determine the MO coefficient simply from one field rotation curve [25], in a way very similarly to determination of AMR or Planar Hall effect coefficients from field rotations [26]. This also directly demonstrates analogy between the magneto-optical and magneto-transport methods.

### III. Theory

The aim of our theoretical analysis is to determine the kinetics of the magnetization vector during the magnetic field sweep and, based on its known orientation at each point of the experimental curves, to evaluate the magnitude of the magneto-optical coefficient. Motion of the magnetization vector is modeled in terms of the local profile of the magnetization free energy density. Its functional $F$ is known from the symmetry considerations (see Eq. (S1) in Supplementary) assuming the lowest terms in magnetization magnitude [41], yet the corresponding constants which appear in the expression are strongly sample-dependent. In the case of YIG, the expected order of magnitude of the anisotropy constants is known [42] and therefore, we can roughly estimate the positions of the easy magnetization directions. The dominant anisotropy in high-quality thin YIG samples on GGG has its origin in the cubic bulk contribution. In thin samples, there is an additional out-of-plane anisotropy (hard direction) due to the stress fields and demagnetization energy which pushes the magnetization towards the sample plane [14,15]. We therefore expect that the projection of easy directions to the crystallographically oriented [111] sample plane is effectively sixfold [see Fig. 3(c)] and the deflection angle of the easy directions from the sample plane is only few degrees and thus will be neglected (see the Supplementary information for more details). We define an effective in-plane anisotropic energy density, assuming that the deviation angle of the magnetization vector from the sample plane is small:

$$\frac{F}{M_S} = K_6 \sin^2 3(\varphi_M - \gamma) - \mu_0 H_{ext} \sin\theta_H \cos(\varphi_H - \varphi_M) \qquad (1)$$

where $M_S$ is the saturation magnetization of the sample, $\mu_0$ is the vacuum permeability, $H_{ext}$ is the external magnetic field magnitude, $\theta_H$ its deflection angle of from the sample plane and $\varphi_H$ its azimuthal orientation with respect to one of the effective in-plane easy axes (assuming $K > 0$). The symbol $\varphi_M$ denotes azimuthal position of the in-plane magnetization vector and $K_6$ is the effective anisotropy constant. $\gamma$ denotes an angle between the plane of incidence and the bisectrix of the magnetization easy axes, resulting from an unintentional rotation of the sample in the experiment.

When interpreting the experimental data, we theoretically simulated the full magneto-optical measurement of the hysteresis by calculating the MO response of the layered structure (nm-thick sample on a 500 μm-thick substrate), considering the optical constants of the participating materials and the symmetry-breaking by the sample's magnetization. Our calculations inherently include all effects related to the light propagation in the media as well as multiple reflections and resulting interferences and as such they reveal the sum of all MO effects which take part in the particular geometry. We consider the refractive index of the thick GGG substrate to be $n_S$ = 1.96 and the YIG permittivity tensor for magnetization oriented along the *x* axis reads:

$$ \qquad (2) $$

where we take $\epsilon_N$ = 6.5+3.4i [42]. Thanks to the cubic symmetry of the YIG crystal, its permittivity tensor for an arbitrary magnetization orientation in the sample's *xy* plane (see the geometry in Fig. 2) is calculated by a proper rotation of Eq. (2) around the *z* axis corresponding to [111] crystallographic direction. We considered the values $Q = 10i \times 10^{-3}$ and $Q_A = 1.1 \times 10^{-3}$ for the simulations of the hysteresis loops for best agreement with measured amplitudes of the MO effects.

The experimental data can be interpreted also in terms of the symmetry of the MO response: they are composed of an even and an odd component whose magnitude remain (invert) upon inversion of the magnetization. The even contribution is related to the Cotton-Mouton effect and is represented by the parameter $Q_A$ in Eq. (2). The even symmetry comes from the fact that the effect is quadratic, i.e. the parameter $Q_A$ is proportional to the square of the magnetization. The odd contribution can be phenomenologically understood as a combination of the longitudinal and the transverse Faraday effect and is described by the parameter $Q$ in Eq. (2), which is linear in magnetization. Note also that the polar Faraday effect does not contribute to the overall MO response of the system because the projection of magnetization to the polar (out-of-plane) direction is negligible.

Considering the incident *s*-polarization ($\beta$ = 0 in our geometry), the polarization rotation $\Delta\beta$ due to the linear MO effect is zero in the transverse magnetization geometry (i.e., for magnetization lying along the *x* direction). It is therefore reasonable to consider phenomenologically that for any arbitrary in-plane orientation of the magnetization, the polarization rotation is solely due to the longitudinal Faraday effect, i.e. it is proportional to the projection of the magnetization to the plane of incidence. We can write for the polarization rotation:

$$ \Delta\beta_{LFE}(\varphi_M, \beta=0) = P^{LFE} \sin\varphi_M \qquad (3) $$

where we defined the effective LFE coefficient $P^{LFE}$. Expressions for other than s-polarizations outside the limit of the small angle of incidence $\vartheta_i$ are not convenient for practical use and therefore are not discussed here; the numerical results, however, will be presented in the text below. In magneto-optical experiments, we always measure the polarization rotation change when magnetization orientation changes. Therefore, we define the LFE amplitude $A_{LFE}$ as:

$$A_{\text{LFE}}(\varphi_1,\varphi_2) = P^{LFE}(\sin\varphi_1 - \sin\varphi_2) \qquad (4)$$

where $\Delta\beta$ is the polarization rotation of incident s-polarization, including all MO and non-magnetic contributions, for a given orientation of magnetization. The difference in each of the square brackets represents the measured change of the MO signal and the subtraction of the parentheses extracts only the linear (odd) LFE component from the MO signal. The angles $\varphi_{1,2}$ denote some well-defined orientations of the magnetization. In our case, it is worth to use positions of the easy axes between which the magnetization jumps during the hysteresis curve measurements.

In contrast to LFE (Eq. 3), CME is sensitive to the angle between magnetization and light polarization direction. Simple relation can be derived, describing the relation between polarization rotation $\Delta\beta_{CME}$ and magnetization position $\varphi_M$ [23]

$$\Delta\beta_{CME}(\varphi_M) = P^{CME} \sin 2(\varphi_M - \beta) \qquad (5)$$

where we defined the effective CME coefficient $P^{CME}$. In the specific case of hysteresis loops where magnetization is switched between two magnetic easy axes, we may define the CME amplitude (see Fig. 4 and Eq. (11) in [40]) to:

$$A_{CME}() = 2P^{CME} \sin\xi \cos 2(\gamma - \beta) \qquad , \qquad (6)$$

where $= \varphi_1 - \varphi_2$ is the angle between the easy axes and $\gamma = (\varphi_1 - \varphi_2)/2$ is the position of their bisectrix. The symmetrization of the brackets ensures that only even MO signal contributes to the amplitude. When the angles $\varphi_{1,2}$ are known, it is possible to extract the MO coefficients using the above expressions.

IV. **Experimental results and discussion**

A. **Hysteresis loops**

Firstly, we focused on studying the MO during external magnetic field sweeps (MO hysteresis loops). In Fig. 3(a) we show an example of the MO hysteresis loop measured close to the normal incidence ($\theta_i = 3°$) and at a large angle of incidence ($\theta_i = 45°$) at 20 K. The character of the hysteresis loop changes significantly when deviating from the normal incidence. Close to the normal incidence, the signal displays an M-shape like loop, typical for the quadratic MO effects [40]. The steps in the M-shape loops generally correspond to a switching of magnetization between two magnetic easy axes [40]. In contrast, for $\theta_i = 45°$ the hysteresis loop gains more complex shape. Besides the M-shape like signal which is still present with virtually unchanged size, there is another square-like component that indicates presence of a signal odd in magnetization, which can be attributed the longitudinal Faraday effect.

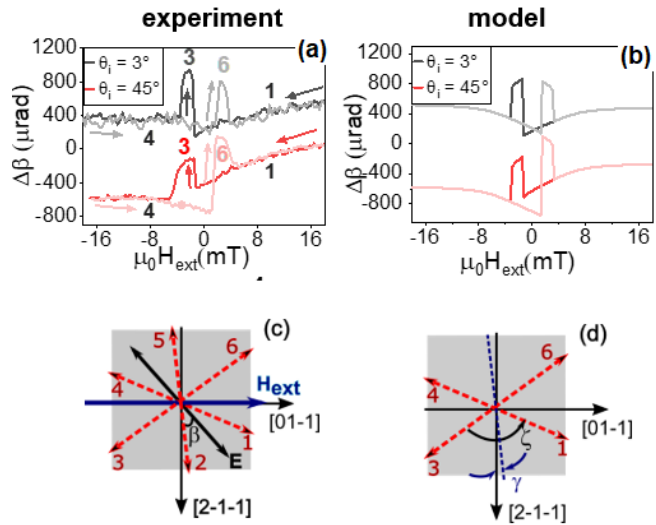

Fig. 3: (a) Rotation of polarization plane $\Delta\beta$ as a function of the external magnetic field $H_{ext}$ measured for two angles of incidence $\theta_i = 3°$ and 45°, at temperature of 20 K and photon energy of 3.1 eV. The data were vertically shifted for clarity. The complex M-shape-like hysteresis is a clear signature of magnetization being switched between magnetic easy axes. (b) Simulation of the MO signal by means of the analytical model. Based on our model, we identified 3 equivalent easy axes (c) and extracted their mutual angle $\zeta = 120°$ and position of their bisectrice $\gamma = 6°$. (d) The abrupt

changes in magneto-optical signals in (a) and (b) correspond to jumps of the magnetization between the easy axes 1, 3 and 4 (4, 6 and 1) for the magnetic field sweep from the positive (negative) field, as schematically indicated in (a).

In order to understand the nature of the magnetization motion, we used the theoretical approach described in the Theory section to model the observed signals: we consider six effective in-plane easy directions for magnetization (see Eq. (1)) and we numerically modelled the MO response considering the parameters of the experiment. We used four fitting parameters: two of them are related to the amplitude of the MO effects (even and odd) and two describe the magnetic anisotropy. The best agreement with the experimental data appears for the values Q = 10i×10$^{-3}$, $Q_A$ = 1.1×10$^{-3}$, $\gamma$ = 6° and $K_6$ = 61 J/$m^3$ . As an output of our model, the correct shape of the MO loops is obtained, as shown in Fig 3(b). The magnetic anisotropy utilized by the model is schematically depicted in Fig. 3(c), with a definition of the magnetic easy axes position given in Fig. 3(d). Note that the estimated magnetic anisotropy corresponds with the SQUID measurement [Fig. 1(d)], the diagonal orientation (denoted as "C") being the closest to the position of one of the easy axes. The motion of magnetization **M** in the external magnetic field gained from our model is schematically indicated by the arrows in Fig. 3(a). For the large positive external field **H**$_{ext}$, **M** is oriented along the field, close to the easy axis (EA) labelled as "1". While decreasing **H**$_{ext}$, **M** is slowly rotated towards the direction of EA "1". When **H**$_{ext}$ of the opposite polarity and magnitude exceeding the value of coercive field H$_c$ is applied, **M** switches directly to the EA "3". Further increase of the negative **H**$_{ext}$ leads to another switching, this time only by 60° to EA "4", until, finally, **M** is again oriented in the direction of **H**$_{ext}$. A symmetrical process takes place in the second branch of the hysteresis loop. Note that the same magnetization switching occurs independently of the angle of incidence, though for larger $\theta_i$ the shape of the loop is distorted by the presence of the linear contribution to the MO signal.

The macrospin simulations confirmed that the full magnetization trajectory extracted from our magneto-optical signals corresponds to the realistic magnetic anisotropy constants for thin YIG films (see Section 2 in Supplementary). The model allows for certain ambiguity in its parameters since we do not have access to the out-of-plane components of magnetization motion during the switching, to compare them with the experimental data. The model thus cannot be used reliably for obtaining all magnetic anisotropy constants of the material without support of a complementary experimental method. However, it provides a useful tool for prediction of the behavior of the magneto-optical effects, as shall be shown further on.

To quantify the contributions of the linear and quadratic MO effect, we recorded the MO hysteresis loops for different angles of initial light polarization β. The quadratic CME and the linear LFE contributions to the overall MO signals were obtained by symmetrization and anitsymmetrization of the hysteresis loops, as indicated in Fig. 4(a) and 4(b) and Eqs. (4) and (6), for the two angles of incidence $\theta_i$ = 3° and 45°, respectively. Note that after the separation, the signal indeed splits into the square-shape hysteresis loop

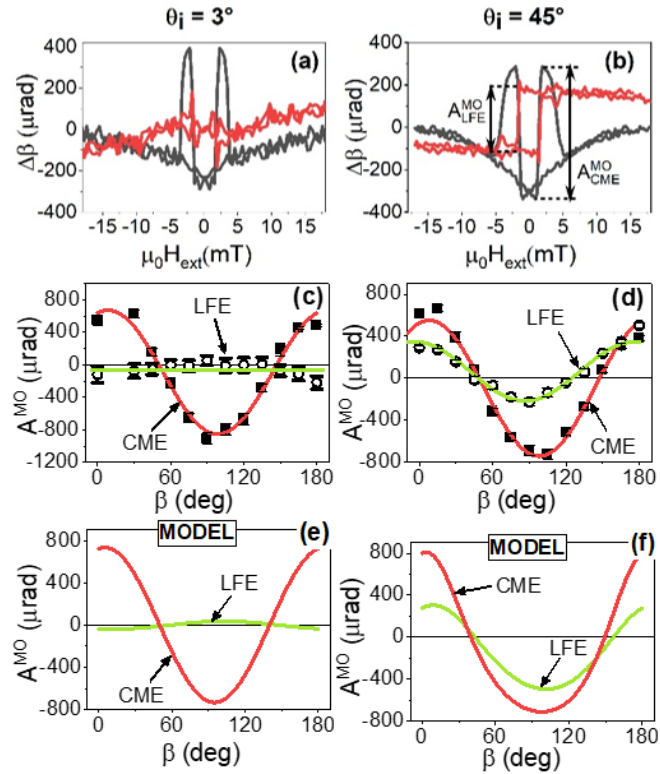

typical for linear MO signals, and the characteristic M-shape loop of the quadratic magnetooptics.

Fig. 4: Analysis of magneto-optical (MO) signals at an angle of incidence $\theta_i$ = 3° (left column) and $\theta_i$ = 45° (right column). For extracting the MO effects odd and even in magnetization, the signals were decomposed to symmetrical (black line) and antisymmetrical (red line) components with respect to $H_{ext}$. An example of results of this procedure is shown on for MO signals measured for angles of incidence $\theta_i$ = 3° (a) and $\theta_i$ = 45° (b) using β = 0°. The original data are those from Fig. 3. The amplitude of the particular MO effect $A_{MO}^{LFE}$ and $A_{MO}^{CME}$ for each polarization angle β was determined from the size of the „jumps" in hysteresis loops, as indicated in (b). The corresponding polarization dependencies of LFE and CME are shown for $\theta_i$ = 3° (c) and $\theta_i$ = 45° (d). Points stand for the measured data. Green line corresponds to fit to Eq. (3), with amplitude $P^{LFE}$ = (310 ± 20) rad, assuming the switching takes place between the easy axes separated by $\phi_1 - \phi_2$ = 120°. Red line is a fit to Eq. (5), where $P^{CME}$ = (450 ± 30) rad for $\theta_i$ = 3°, and $P^{CME}$ = (320 ± 20) rad for $\theta_i$ = 45°. Further comparison with the analytical model for polarization dependence of MO signals at the two angles of incidence for $\theta_i$ = 3° (e) and $\theta_i$ = 45° (f) show an excellent agreement, confirming validity of the model. Parameters of the model are the same as for Fig. 3

For further analysis we need to determine amplitudes of the MO effects attributed to the particular magnetization switching process. The amplitude of the even component $A_{CME}$ is taken from the first 120° magnetization switching, as indicated in Fig. 4(b). The amplitude of the odd component , is obtained from the same 120°switching as the size of the square-shape signal, i.e. $\varphi_1 = -\varphi_2 = 60°$ in Eq. (5). This method also eliminates potential contributions from the paramagnetic GGG substrate, where no step-like hysteretic behavior is expected.

The resulting amplitudes of the separated MO signals are shown as a function of the light polarization in Fig. 4(c) and (d) for the angles of $\theta_i = 3°$ and 45°. Points in the graphs indicate the values extracted from the experiments, lines are fits by Eqs. (3) and (5), from which the values of the effective MO coefficients $P^{CME} = (320 \pm 20)$ rad and $P^{LFE} = (310 \pm 25)$ rad were extracted for the 45° incidence angle, and $P^{CME} = (450 \pm 30)$ rad for the near-normal incidence. Note that even for $\theta_i = 45°$, which is optimal for observation of the longitudinal Faraday effect, the strength of the quadratic CME exceeds that of the linear LFE, and reaches the values known from Heusler alloys, which are among the highest observed so far [28].

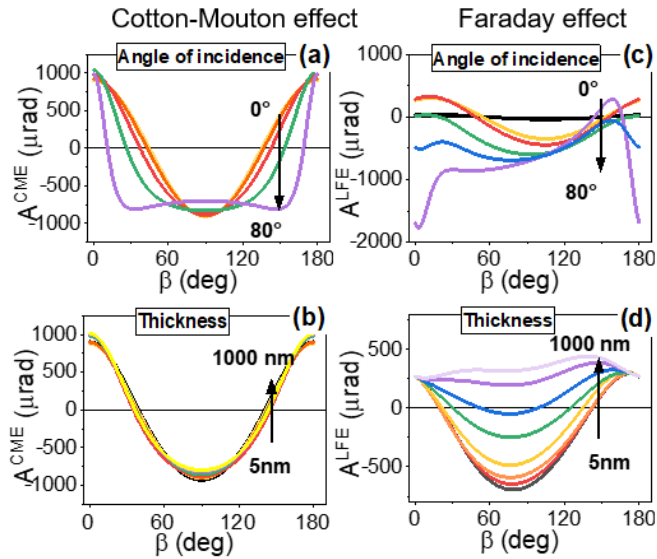

Fig. 5: Amplitudes of CME (left column) and LFE (right column) effect as a function of initial polarization angle β, extracted from the hysteresis loops obtained from the analytical model. The amplitudes $A^{CME}$ and $A^{LFE}$ are obtained by the same method as in Fig. 4.
Polarization dependence of (a) CME and (b) LFE effects for various angles of incidence $\theta_i$ and fixed sample thickness of 50 nm. The angles of incidence of $\theta_i = 0,30,40,60$ and 80° are shown, the arrow indicates increase of $\theta_i$. Clearly, the LFE is very strongly angle-dependent, while CME is much less affected.
The same feature can be observed for changing of the sample thickness *d*. While the polarization dependence of CME (c) does not depend on the sample thickness, LFE (d) can vary significantly. The thicknesses of d = 5,10,20,50,100,200,500 and 1000 nm are displayed for the angle of incidence $\theta_i = 45°$

The observed polarization dependence of MO signal amplitudes can be understood in terms of our analytical model. Keeping all the input parameters of the model fixed, we calculated polarization dependencies of the individual MO amplitudes extracted from the modelled MO hysteresis loops. Resulting dependencies are presented Fig. 4.(c) and 4(d) for $\theta_i$ = 3° and 45°, respectively. The theoretical curves follow the experimental data very well even for the LFE signal close to the normal incidence, which proofs the validity of our analytical approach. We can therefore extend the predictions of the model to conditions that are not easy to systematically change in experiments, particularly the dependence on the angle incidence and the sample thickness. In Fig. 5 we illustrate these dependencies separately for the quadratic CME (graphs in the left column) and linear LFE (right column). The material parameters for each curve are set such that $P^{CME}$ = 450 rad at near-normal incidence and $P^{LFE}$ = 310 rad for the 45° angle of incidence. We also consider sample rotation angle γ=0° for simplicity. We did not simulate the full magnetization dynamics for the purpose of Fig. 5 but we rather used a simplified scheme where we consider 120° magnetization change for CME and LFE. Remarkably, there is a significant difference in how the polarization dependence of the linear and the quadratic MO effect is affected by changing both the sample thickness and the angle of incidence. Non-intuitively, the strengths CME is only weakly affected by both these parameters. The shape of polarization dependence is modified for large angles of incidence $\theta_I$ [Fig. 5(a)] but the maximum value of $A_{CME}$ remains virtually unchanged. The sample thickness has almost no effect on the CME signals [Fig. 5(b)]. In contrast, the linear MO signals are drastically modified by both these parameters. As expected, the linear MO effect decreases for smaller angles of incidence [Fig. 5(c)], eventually disappearing at normal incidence. However, not only the magnitude but also the shape of the polarization dependence is affected. This complex behavior of LFE results from interferences: while CME in our case results only from magnetic linear dichroism (i.e., the difference of absorption coefficient of the two orthogonal optical polarization eigenmodes), LFE is a consequence of birefrigence of elliptically polarized eigenmodes (i.e., the difference of the corresponding effective refractive indexes). As a consequence, extrema of the MO response appear at resonances. We observe in Fig. 5(d) only a monotonous trend of the curve shaping with the increasing sample thickness because of a relatively large sample absorption at the used wavelength which prevents multiple wave roundtrips inside the YIG layer even at the position of the first resonance. The shape of the LFE response therefore relaxes from the zero-order resonance (small sample thickness) up to no multiple reflections for large sample thickness and becomes saturated. This complex modification of LFE response makes it difficult to optimize the sample thickness for magnetometry measurements, and using the quadratic CME therefore provides a significant advantage.

It is important to stress that our model is independent of the magnetic anisotropy of the particular sample. The conclusions drawn from the model are, therefore, universal for a series of samples with identical bulk magnetic properties.

### B. ROT-MOKE measurements

In order to further investigate the nature of the Cotton-Mouton effect, we performed ROT-MOKE measurements [24-25] close to normal incidence geometry to eliminate the linear MO contribution to the signal. The ROT-MOKE method provides a more efficient and sensitive tool for extracting the MO coefficients, without necessity of modification of the initial light polarization orientation.

Examples of the as-measured ROT-MOKE signals are shown as open symbols in Fig. 6(a) and 6(b) for low temperature (T= 10K) and room temperature, respectively. As we are interested in even MO signals only, the small contribution of the linear MO effect was removed by symmetrization of the curve with respect to the angle of the external field $\phi_H$ [solid symbols in Fig. 6(a) and 6(b)]. The symmetrized curves display a clear harmonic behavior, which indicates that the field of 205 mT was large enough to saturate magnetization, which follows exactly the direction $\phi_H$. We were therefore able to fit the data to equation (4) (red line) with angle of magnetization equal to the angle of $\mathbf{H_{ext}}$ ( $\phi_H = \phi_M$ ). From the fits we obtain the MO coefficient $P^{CME}$= (450±40) rad at 10 K, which is in excellent agreement with the value extracted from the hysteresis loops. However, the value of $P^{CME}$ coefficient decreases to $P^{CME}$= (230±20) when heating the sample to room temperature. To understand this change, we measured the temperature dependence of $P^{CME}$. Generally, since the CME is of the second order in magnetization, scaling of $P^{CME}$ with a square of saturation magnetization $M_s$ is expected [22,31]. As illustrated in Fig. 6(c), the good correlation between these two quantities confirms the intrinsic magnetic origin of the CME effect.

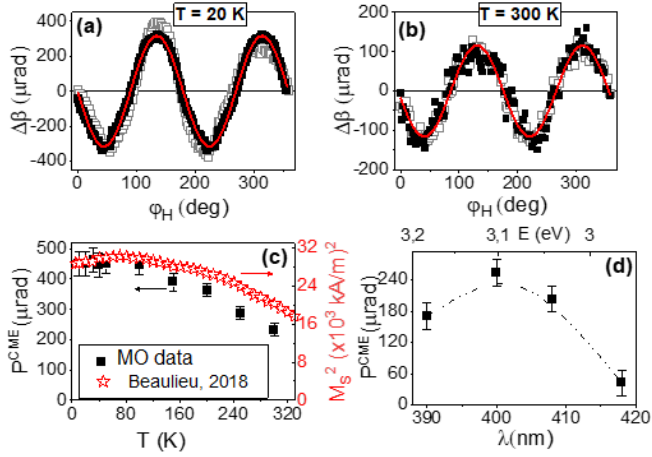

Fig. 6: Rotation of polarization plane Δβ as a function of the direction $\varphi_H$ of the external magnetic field of a fixed magnitude $\mu_0 H_{ext}$ = 205 mT, measured at 20 K (a) and at room temperature (b). Polarization was set to β =0°. Open squares indicate the as measured data, full squares are symmetrized in $\varphi_H$ to remove linear magneto-optical effects. Red line is a fit to Eq. (5) for $\varphi_H = \varphi_M$ as $H_{ext}$ is large enough to saturate magnetization of the sample. Magneto-optical coefficient for Cotton-Mouton effect obtained from the fits are $P^{CME}$ = 450 rad at 20 K and $P^{CME}$ = 230 rad at 300 K. The decrease of $P^{CME}$ with increasing temperature is well correlated with the reduction of square of the saturation magnetization $M_S^2$, values of which were taken from reference Ref. 44 and converted to SI units (c). The spectral dependence of $P^{CME}$ measured at room temperature (d) clearly shows a peak at around 3.1 eV.

The physical origin of the intrinsic CME effect can be unveiled by its spectral dependence. For this purpose, we extracted $P^{CME}$ coefficient from the room-temperature ROT-MOKE data measured at several wavelengths. The obtained spectrum of $P^{CME}$ is presented in Fig. 6(d). The maximum of CME occurs at around λ = 400 nm (3.1 eV), and its amplitude drops rapidly when the laser is detuned from the central wavelength by more than 10 nm. The sharp increase of the CME response around 3.05-3.1 eV corresponds energetically to transitions O-2p to Fe-3d band states of YIG [45]. Giant Zeeman shift of this transition level was recently reported in a 50 nm thin [111] YIG film [45]., which is very similar to the sample studied in our work. Its origin was attributed to the combination of strong exchange interaction of Fe-3d orbitals and the effect of spin-orbit coupling on the Fe-3d bands. Similar combined act of the exchange of magnetic ions and spin-orbit coupled valence bands is known from diluted magnetic semiconductors [28,46], the systems that are typical for their strong quadratic MO response with a significant peak on the Zeeman-split energy level [46]. Analogously, strong quadratic response of the YIG thin layers can be expected [46].

Apart from the intrinsic MO effect, impurity states can significantly influence the MO response of thin films. Lattice defects are known to occur during growth of the very thin YIG layers, particularly due to the migration of $Fe^{3+}$ and $Gd^{3+}$ ions across the interface during the post-growth annealing [14,15,18]. Similarly, gadolinium doping can be responsible for the decrease in saturation magnetization of the PLD-grown thin

YIG layers [17]. However, it affects mostly the interfacial layer of a few nanometers, which orders antiferromagnetically, reducing the magneto-optical response of the layer [47], and cannot thus be responsible for the origin of the observed Cotton-Mouton effect. In fact, previous works based on the quadratic MO response of YIG [20-21] were always performed in a spectral region close to 400 nm, even though thin films of various thickness, prepared by different methods and presumably containing different level of impurities, were studied. Though the choice of the wavelength was not performed systematically in these works and the amplitude of CME was not evaluated, the wavelengths always lay close to the optimum value that was identified from our experiments. We are thus led to a conclusion that the observed strong CME response is very likely intrinsic to any YIG thin layer, and it is not related to unintentional doping or any type of defects. Therefore, the MO effect seems to be universally applicable.

## V. Conclusions

In summary, we have shown the presence of a strong Cotton-Mouton effect in a 50-nm thick epitaxial layer of YIG in the spectral region close to 400 nm. We measured both magneto-optical hysteresis loops and ROT-MOKE data that enabled us to extract the values of CME coefficient. The maximum $P^{CME}$= 450 rad obtained for our YIG layer is comparable to the giant quadratic magneto-optical response of Heusler alloys [27] or ferromagnetic semiconductor GaMnAs [28]. Spectral and temperature dependence both indicate an intrinsic origin of the effect, which demonstrates its universal applicability for the magneto-optical magnetometry. This functionality of the MO experiment was demonstrated in determining cubic magnetocrystalline anisotropy of the thin YIG film, which is the dominant magnetic anisotropy in the sample.

The measured signals were further analyzed using an analytical model based on a calculation of the overall optical and magneto-optical response of the thin YIG layer on GGG substrate. The model enables to predict properties of longitudinal Faraday and Cotton-Mouton effect for variable sample thicknesses and angles of incidence, which are parameters crucial for many thin-film experiments. The calculation revealed that while LFE varies strongly both with angle of incidence and sample thickness, CME has a comparable magnitude but much weaker sensitivity to both the studied parameters. Therefore, using the quadratic CME provides an advantage against the linear LFE, particularly when normal incidence is dictated by the experiment geometry, which is the case for most of the opto-spintronics experiments. Our combined theoretical and experimental approach enables to optimize the condition for the experiment in terms of choice of proper light source or measurement geometry, which can lead to a significant increase of signal-to-noise ratio and sensitivity in opto-spintronics experiments.

**Acknowledgmements:**

E.S. and T.O. contributed equally to the work.

The authors would like to acknowledge fruitful discussions with Dr. Jaroslav Hamrle and Dr. Eva Jakubisová. This work was supported in part by the INTER-COST grant no. LTC20026 and by the EU FET Open RIA grant no. 766566. We also acknowledge CzechNanoLab project LM2018110 funded by MEYS CR for the financial support of the measurements at LNSM Research Infrastructure as well as the German research foundation (SFB TRR173 Spin+X 268565370 - projects A01 and B02).

# Giant quadratic magneto-optical response of ultra-thin YIG films for sensitive magnetometric experiments: Supplementary material

E. Schmoranzerová[1]*, T. Ostatnický[1]*, J. Kimák[1], D. Kriegner[2,3], R. Schlitz[3], H. Reichlová[2,3], A. Baďura[1], Z. Šobáň[2], M. Münzenberg[4], G. Jakob[5], E.-J. Guo[5], M. Kläui[5], P. Němec[1]

[1]Faculty of Mathematics and Physics, Charles University, Prague, 12116, Czech Republic

[2]Institute of Physics ASCR v.v.i , Prague, 162 53, Czech Republic

[3]Technical University Dresden, 01062 Dresden, Germany

[4]Institute of Physics, Ernst-Moritz-Arndt University, 17489, Greifswald, Germany

[5]Institute of Physics, Johannes Gutenberg University Mainz, 55099 Mainz, Germany

1. **Magnetic characterization**

Magnetic properties of the samples were characterized by SQUID magnetometry. SQUID magnetic hysteresis loops, detected at several ambient temperatures for a fixed direction of the external magnetic field along [2-11], are shown in Fig. S1(a). As expected [s1], the values of magnetic moment in saturation increase with decreasing temperatures. Simultaneously, the coercive field slightly increases but remains well below 20 Oe in the whole temperature range.

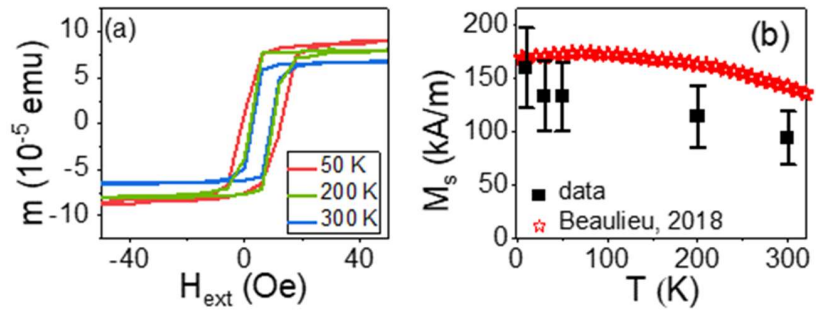

Fig. S1: SQUID magnetometry at variable temperatures. (a) Magnetic hysteresis loops measured by SQUID magnetometry in direction [2-1-1] at several ambient temperatures. As expected, the saturation magnetization $M_s$ increases when temperature is decreased, which is accompanied by a slight increase in coercive field. The temperature dependence is highlighted in (b). The values of $M_s$ were recalculated to CGS units for comparison with values in Ref. 44

The values of saturation magnetization $M_s$ can be extracted from the SQUID hysteresis loops by recalculating to the volume of the magnetic layer. However, this parameter is burdened with a relatively large error in our experiment. The thickness of the crystalline YIG layer of d = (46.0 ±2.4) nm is known from the XRD experiment. The sample area is estimated to S = (23 ±3)x10$^{-2}$ cm$^2$ , with the error resulting from an irregular shape of the sample. Furthermore, the SQUID hysteresis loops were affected by a strong paramagnetic background of the GGG substrate, which also increased the error in estimation of $M_s$. These issues do not allow us to determine precisely the detailed temperature dependence of $M_s$, required for comparison with the magneto-optical experiment (see Fig. 6 in main text). Instead, we measured the values of $M_s$ only at several temperatures to verify the general trend of the dependence. In Fig. S1(b), the measured values of saturation magnetization $M_s$ are compared with the published results, obtained on a nominally similar sample (Ref. 44). Clearly, the data follow a similar trend and the values of $M_s$ are comparable in size, which justifies utilization of the published results for comparison with our magneto-optical data in the main text.

## 2. Tracing the magnetization trajectory in the measurement of the hysteresis loops

The knowledge of the magnetization vector path in the [111] oriented YIG thin layer during the measurement of the hysteresis is essential in interpretation of the experimental data and determining the magneto-optical response coefficients. We observe two distinct points in Fig. 3(a) in the main text in each of the branches of the hysteresis loops where the magneto-optical signal abruptly changes when smoothly changing the external magnetic field magnitude. These points represent a rapid change (jump in the following) of the magnetization vector from the vicinity of one magnetization easy axis towards another one. For the interpretation of the data, it is then essential to know between which directions the magnetization jump takes place.

The discussion of [111] oriented layers of cubic materials is not straightforward like in [001] oriented samples. In the latter case, the demagnetizing field and the uniaxial out-of-plane anisotropy usually push the magnetization vector to the plane of the sample where there are at most four easy directions for the magnetization due to the cubic symmetry of the material. In the case of [111] orientation, on the other hand, all the easy axes lie out of the sample plane. The demagnetizing field can be strong enough to tilt the easy axes to the close vicinity of the sample plane, however, they always provide a nonzero out-of-plane component. The projection of the three easy axes directions to the sample plane reveals a sixfold symmetry in the contrast to [001] oriented layers with a fourfold symmetry. This behavior is illustrated in Fig. S2 where we plot the magnetization free energy functional $F$ as a function of the polar ($\theta_M$) and azimuthal ($\varphi_M$) angle. We consider the form of the functional including first- and second-order cubic terms [s3,s4] and we define the polar angle $\theta$ with respect to the crystallographic axis [111] and the azimuthal angle $\varphi = 0$ in the direction $[2\bar{1}\bar{1}]$ with an appropriate index referring to the magnetization position (index $M$) or the direction of the external magnetic field (index $H$). The resulting functional takes the form (in the SI units)

$$F = -\mu_0 HM[\sin\theta_M \sin\theta_H + \cos\theta_M \cos\theta_H \cos(\varphi_H - \varphi_M)] + \left(\tfrac{1}{2}\mu_0 M^2 - K_\mathrm{u}\right)\sin^2\theta_M$$
$$+ \frac{K_{c1}}{12}\left[7\cos^4\theta_M - 8\cos^2\theta_M + 4 - 4\sqrt{2}\cos^3\theta_M \sin\theta_M \cos 3\varphi_M\right]$$
$$+ \frac{K_{c2}}{108}\big[-24\cos^6\theta_M + 45\cos^4\theta_M - 24\cos^2\theta_M + 4$$
$$- 2\sqrt{2}\cos^3\theta_M \sin\theta_M(5\cos^2\theta_M - 2)\cos 3\varphi_M + \cos^6\theta_M \cos 6\varphi_M\big],$$

(S1)

where $\mu_0$ is the vacuum permeability and we consider the following values of constants [s5] of a bulk material: magnetization $M$ = 196 kA/m, first-order cubic anisotropy constant $K_{c1}$ = −2480 J/m$^3$, second-order cubic anisotropy constant $K_{c2}$ = −118 J/m$^3$. The external magnetic field is set to zero for the purpose of Fig. S2: $\mu_0 H$ = 0 mT. The uniaxial out-of-plane anisotropy is set to compensate the effect of the demagnetization $K_u$ = 24 kJ/m$^3$ in Fig. S2(a) while we consider $K_u$ = 0 J/m$^3$ in Fig. S2(b). We clearly observe that the demagnetizing field causes the drag of the energy density minima towards the sample plane, leading to a deviation angle (relative to the sample plane) of only a few degrees. At the same time, it weakens the resulting total magnetic anisotropy.

Our analysis of the motion of the magnetization vector in the external field is based on an estimation of the magnetic anisotropic constants of the sample. These constants are determined from positions of the jumps in the hystereses in Fig. 1(c)-(d) in the main text, which were measured using the magnetic field inclination angle with respect to the sample plane $\theta_H$ = 0° and $\theta_H$ = 45°. The value of the low-temperature magnetization of our sample is $M$ = 174 kA/m [2186 G in cgs units, see Fig. S1(b)] and, according to the bulk values, we set $K_{c2} = K_{c1}/21$ [s5]. The out-of-plane uniaxial anisotropy is, however, uncertain and strongly sample-specific (see e.g., Refs. 18 and 40 of the main text). We therefore performed a fit of the experimental data considering the value $K_u$=0, resulting in the values $K_{c1}$=−4.68 kJ/m$^3$ ($k_{c1} = K_{c1}/2\pi M$ = −540 Oe). This number is higher compared to the bulk value reported in Ref. S5. On the other hand, other published low-temperature values are even higher [s6] or reveal a clear tendency [s1] that the cubic anisotropy constant could be significantly higher than the bulk value.

As shown in Fig. S2(b), the magnetization free energy density reveals a very narrow valley. The magnetization vector is constrained in the polar direction by the demagnetizing field while modulation of its effective potential is weak in the azimuthal direction. We may therefore regard its motion in the azimuthal direction in weak magnetic fields as effectively one-dimensional, similarly to the [001] oriented layers. The effective one-dimensional free energy density is then calculated from Eq. (S2):

$$F_{\text{eff}}(\varphi_M) = \min_{\theta_M} F(\theta_M, \varphi_M) \tag{S2}$$

where $\min_{\theta_M}$ denotes the minimum value with respect to angle $\theta_M$.

We plot the effective free energy density for the fitted value of the cubic anisotropy $K_{c1}$= −4.68 kJ/m$^3$ in Fig. S3. The cosine-like curve can be fitted by an effective one-dimensional free energy functional:

$$F_{\text{inplane}} = K_6 \sin^2 3\varphi_M \tag{S3}$$

with the effective $K_6$ anisotropy constant defined in accordance to the cubic in-plane anisotropy of [001] oriented layers. Figure S3 shows the comparison of the effective free energy density as defined by Eq. (S2) (black solid line) and the fitted Eq. (S3) (blue dashed line). The two curves coincide and therefore we may regard the system as effectively in-plane with the fitted value of the sixfold anisotropy $K_6$ = 245 J/m³ ($k_6$ = $K_6/2\pi M$ = 28 Oe).

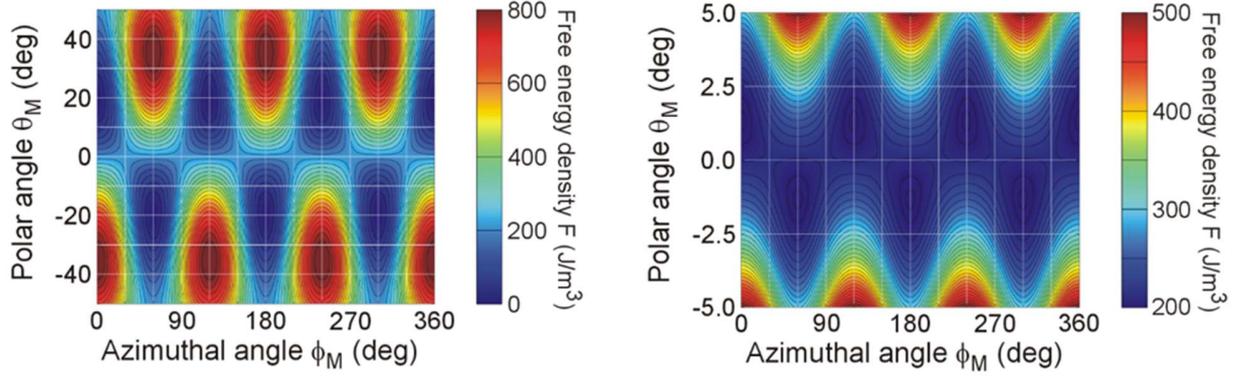

Fig. S2: Magnetization free energy density in [111] oriented YIG sample considering parameters of a bulk sample [s3]: M=196 kA/m, $K_{c1}$=- 2480 J/m3, $K_{c2}$= -118 J/m3. The plots represent the bulk material with (a) the demagnetizing field exactly compensating the out-of-plane anisotropy and (b) no effect of the demagnetization. Note that there is a different y-scale in parts (a) and (b).

As noted above, the strength of the uniaxial out-of-plane anisotropy is not known and the value $K_u$=0 kJ/m³ has been used in the fitting procedure. It is therefore necessary to verify the robustness of the modelled cubic anisotropic constant against nonzero $K_u$. Using, for example, the value $K_u$=10.4 kJ/m³ ($k_u$ = $K_u/2\pi M$=1.2 kOe, approx. one half of the strength of demagnetization), we fit the cubic anisotropy constant $K_{c1}$=−3.48 kJ/m³ (−400 Oe). The results are also presented in Fig. S3 (red solid line). Clearly, the effective in-plane free energy density changes as compared to the situation with $K_u$=0 but the difference is only of the order of 10%. We may therefore conclude that the exact knowledge of the out-of-plane anisotropy is not necessary for the analysis of the in-plane magnetization dynamics and we may consider $K_u$=0 in the forthcoming calculations.

Besides the cubic anisotropy, an additional stress-induced uniaxial anisotropy oriented in the plane of the sample is known to occur [s7]. We took this fact into consideration, performing the fit of the experimental data by taking the strength and the orientation of the anisotropy field as free parameters. Our results show that the effect of any eventual in-plane stress is negligible. This agrees with the reciprocal space map measurements, presented in Fig. 1(c) of the main text.

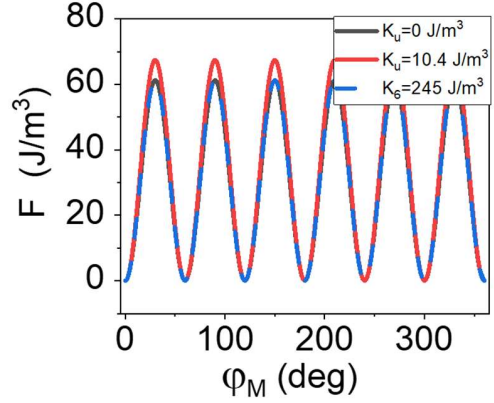

Fig. S3: Effective free energy density F as a function of in-plane (azimuthal) magnetization angle $\phi_M$, calculated from Eq. (S1) by considering the parameters of Fig. (S2), and the out-of-plane uniaxial anisotropy of $K_U$= 0 J/m$^3$ (black solid line) and $K_U$= 10.4 J/m$^3$ (red solid line). The blue dashed line represents a fit by a sixfold in-plane anisotropy using Eq. (S3), giving the effective parameter $K_6$= 245 J/m$^3$

As a result of the analysis of the sample magnetic anisotropies, we can plot the trajectory of the magnetization vector during the measurement of the hysteresis loops. We consider for this purpose no uniaxial out-of-plane anisotropy ($K_u$=0) and the appropriate values of the cubic anisotropic constants $K_{c1}$ = −4.68 kJ/m$^3$ and $K_{c2}$ = $K_{c1}$/21. The resulting trajectory is compared with the experimental hysteresis curves shown in Fig. S4(a). Here the points where the magnetization rapidly changes its orientation are marked by letters A-J. The situation is schematically depicted in Fig. S4(b), where the in-plane projections of the easy directions (red lines) and projections of the magnetization vector (blue arrows) at the positions marked in Fig. S4(a) are presented. The green arrows then depict the sense of the magnetization motion and its speed: solid lines stand for slow rotation of the magnetization vector while the dotted lines mean rapid changes (jumps). The magnetization motion naturally depends on the size of the uniaxial anisotropy that changes the free energy density F (see Fig. S2). In Figs. S4(c)-(d) we plot the dependency of the azimuthal and polar angles of the magnetization vector on the external field for one branch of the hysteresis. The dependencies are depicted for both cases of the zero and nonzero out-of-plane uniaxial anisotropy and also for the effective model of the sixfold in-plane anisotropy (dashed curve). As we may expect, the curves coincide in the plot of the azimuthal angle while they reveal more pronounced differences in the polar angle. The tracing of the exact magnetization path in the out-of plane direction is not, however, the subject of our discussion, since it does not reflect significantly in our measurement. Instead, the graphs help us to confirm the correctness of the data analysis described below.

The plots in Fig. S4 show two jumps of the magnetization vector for each of the branches of the hysteresis loops between the points B, C and D, E for one branch of the hysteresis loop, and G, H and I, J for the other branch. We may observe in Fig. S4(c) that the in-plane orientation of the magnetization changes by 120°(B→C) and 45°(D→E). The change of the out-of-plane orientation is zero in the first case while it is nonzero in the latter, as apparent from Fig. S4(d). While we could neglect the small deflection of the magnetization from the sample plane in the analysis of the cubic in-plane anisotropy, it plays a significant role in magneto-optical (MO) experiments, as it generates a contribution due to the polar MO effect. The magnetization jumps D→E is the case in which the deflection angle rapidly changes and therefore the change of the MO response during this jump is composed of both the in-plane and out-of-plane components which cannot be further separated. The magnetization out-of-plane component stays conserved, on the contrary, during the B→C jump which is then the right point for extraction of the in-plane MO effect amplitudes, as depicted in Fig. 4(b) and explained in the main text. Note also that the magnetization change between any pair of labelled points in Fig. S4 reveals a nonzero out-of-plane component change except for the B→C jump which then turns out to be the only choice for the experimental determination of the magneto-optical observables.

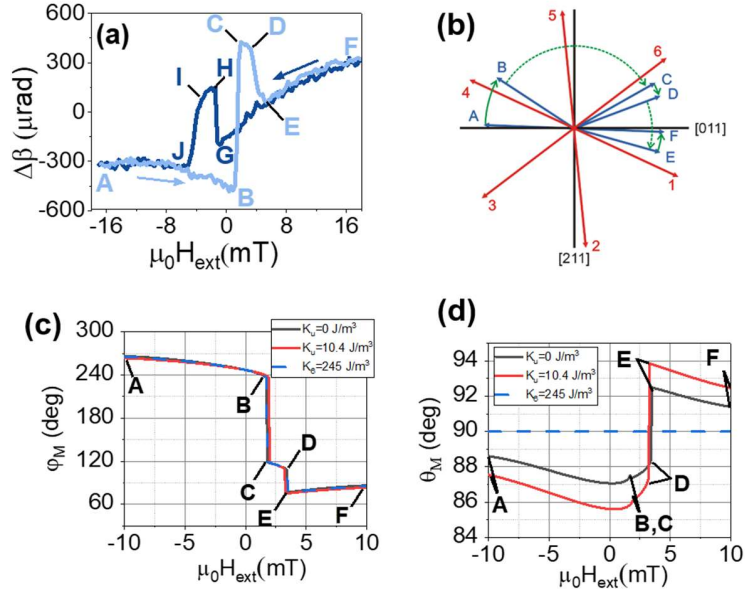

Fig. S4: Tracing of the magnetization vector path when external magnetic is applied at $\theta_H$ = 45 deg in the out-of-plane direction: (a) Hysteresis loop obtained from the magneto-optical experiment with highlighted important points A-F and F-J for the two branches of the loop, respectively. (b) Schematic representation of easy axes (red lines) and positions of magnetization during the switching process (blue lines). Green solid arrows represent slow motion of magnetization, while dashed arrows stand for „steps" in magnetization orientation. (c) Azimuthal and (d) polar angle dependency of magnetization orientation on external field magnitude, modelled for different anisotropy constants $K_u$ and $K_6$. The labelled points match the points in (a) and (b), respectively.